\documentclass[11pt]{article}
\usepackage[utf8]{inputenc}
\usepackage{amsmath,fullpage,graphicx,color,amsfonts,url,multirow,amssymb,tabularx}
\usepackage{natbib}

\newcommand\independent{\protect\mathpalette{\protect\independenT}{\perp}}
\def\independenT#1#2{\mathrel{\rlap{$#1#2$}\mkern2mu{#1#2}}}

\newcolumntype{L}[1]{>{\raggedright\arraybackslash}p{#1}}
\newcolumntype{C}[1]{>{\centering\arraybackslash}p{#1}}
\newcolumntype{R}[1]{>{\raggedleft\arraybackslash}p{#1}}

\title{A causal inference framework for cancer cluster investigations  using publicly available data}

\author{Rachel C. Nethery$^1$, Yue Yang$^1$, Anna J. Brown$^2$, Francesca Dominici$^1$\\
{\small
$^1$Department of Biostatistics, Harvard T.H. Chan School of Public Health}\\
{\small
$^2$Department of Statistics, University of Chicago}
}
\date{}

\begin{document}
\maketitle

\begin{abstract}
Often, a community becomes alarmed when high rates of cancer are noticed, and residents suspect that the cancer cases could be caused by a known source of hazard. In response, the CDC recommends that departments of health perform a standardized incidence ratio (SIR) analysis to determine whether the observed cancer incidence is higher than expected. This approach has several limitations that are well documented in the literature. In this paper we propose a novel causal inference framework for cancer cluster investigations, rooted in the potential outcomes framework. Assuming that a source of hazard representing a potential cause of increased cancer rates in the community is identified a priori, we introduce a new estimand called the causal SIR (cSIR). The cSIR is a ratio defined as the expected cancer incidence in the exposed population divided by the expected cancer incidence for the same population under the (counterfactual) scenario of no exposure. To estimate the cSIR we need to overcome two main challenges: 1) identify unexposed populations that are as similar as possible to the exposed one to inform estimation of the expected cancer incidence under the  counterfactual scenario of no exposure, and 2) publicly available data on cancer incidence for these unexposed populations are often available at a much higher level of spatial aggregation (e.g. county) than what is desired (e.g. census block group). We overcome the first challenge by relying on matching. We overcome the second challenge by developing a Bayesian hierarchical model that borrows information from other sources to impute cancer incidence at the desired a level of spatial aggregation (e.g. census blocks). In simulations, our statistical approach was shown to provide dramatically improved results, i.e., less bias and better coverage, than the current approach to SIR analyses. We apply our proposed approach to determine whether trichloroethylene vapor exposure has caused increased cancer incidence in Endicott, New York.
\end{abstract}

\section{Introduction}
Across the United States, citizens routinely recognize higher than expected rates of cancer in their community and request that the local health department conduct an investigation, with hopes of identifying a common cause. According to a review by \cite{goodman2012cancer}, at least 2,876 cancer cluster investigations were conducted by health departments in the US between 1990 and 2011, most of which were initiated in response to community alarm at high numbers of cancer cases and fear of a connection between the cancer cases and a known environmental exposure. Shockingly few of these investigations, however, have succeeded in identifying a link between the cancer cases and a common exposure, with the vast majority providing no clear answer to the concerned community.

\subsection{Cancer cluster investigation protocol}
The Centers for Disease Control and Prevention (CDC) has provided a protocol to guide health departments in responding to requests for cancer cluster investigations \citep{cdc2013investigating}. The first challenge in this process is that the definition of a cancer cluster is notoriously vague and contested. The CDC defines a cancer cluster as ``a greater than expected number of cancer cases that occurs within a group of people in a geographic area over a defined period of time'' \citep{cdc2013investigating}.
They recommend that cancer cluster investigations proceed by first performing a standardized incidence ratio (SIR) analysis to determine whether the cancer incidence experienced by the community represents a statistically significant elevation compared to what would be expected. The SIR is estimated as the ratio of the observed cancer incidence to the expected incidence based on background rates, and uncertainties and p-values are computed \citep{sahai1993confidence} to determine whether the SIR significantly exceeds the null value of 1.

If statistical significance is found, then the event constitutes a cancer cluster by their definition. Only if a cancer cluster is confirmed does the CDC recommend that health departments proceed to seek possible environmental causes. If the statistical evidence for a cancer cluster is strong and an epidemiological study to test for relationships between environmental factors and the cancer cases is deemed ``warranted'' and ``feasible'', then the CDC suggests conducting such a study.

Although widely used, the protocol for cancer cluster analyses described above has recently been highly criticized on practical grounds, with objectors pointing to the fact that such analyses rarely lead to definitive identification of the cause(s) of the cluster. With no conclusion about the cause(s) of a cluster, simply the identification of one provides no guidance for the concerned public in removing the cause or preventing future cases. In a review of cancer cluster investigations from 1990-2012 \citep{goodman2012cancer}, it is reported that out of the 428 considered, 72 clusters were confirmed, but only three were linked with hypothesized exposures and merely one revealed a clear cause. The many challenges in cancer cluster investigations have been summarized in recent years \citep{goodman2014cancer}. 


The most prominent statistical limitation is the silent multiple comparison problem, also known as the Texas Sharpshooter problem \citep{bender1995statistical}. This issue arises due to the reliance on observed outcome data to inform the development of the statistical hypothesis, i.e., the hypothesis that a cancer cluster is present in the chosen community and time period is only formed in response to the observation that an unusually high number of cancer cases has occurred. Fundamental statistical principles dictate that, occasionally, the locations of cancer diagnoses will cluster together in space and time due to chance alone, i.e. not due to any common cause. Thus, we would expect, due only to chance, to occasionally encounter what appears to be an unusual excess of cancer cases within a small area. Thus, if we first evaluate the spatial distribution of the cancer cases and draw a boundary around a small area that appears to have a high cancer incidence, and then ask if that area is experiencing a higher cancer incidence than expected by chance, we inflate the probability of finding a false positive. Moreover, we are unable to effectively estimate how many multiple comparisons to adjust for, given the near infinite combination of different possible area boundaries, time periods and diseases that could be assessed (hence the term ``silent'' multiple comparisons). Therefore, even multiple comparisons adjusted p-values can range dramatically and thus are not reliable indicators of peculiarity of the event under study.

There have been attempts to address the silent multiple comparison problem, including the introduction of Bayesian methodology \citep{coory2009bayesian} in which uncertainty arising from the multiple comparison problem can be accounted for through the prior distribution. However, the presence of alternative approaches has failed to produce changes in the way cancer cluster analyses are carried out. In order to completely avoid the multiple comparison problem, it has also been suggested that cancer clusters investigations should abandon statistical analyses entirely \citep{coory2013assessment}. 

\subsection{Causal inference approach to SIR analyses}

We take the position that statistical analyses can provide important insights to cancer cluster investigations; however, in order to produce useful and reliable inference, both the overall protocol and the statistical procedures must be overhauled. In this paper, we propose a number of changes to the cancer cluster investigation protocol in order to situate it within a causal inference framework. The most notable procedural change required is that suspected cause of increased cancer incidence in the community, i.e. a putative source of hazard, must be identified prior to any statistical analyses. The CDC's current approach is statistically backwards in that it tests for elevated cancer risk in a community prior to identifying putative sources of hazard that could be responsible for such elevation. Because no specific source(s) of exposure are postulated prior to analysis, the geographic region, time period, and disease types used to formulate the statistical hypothesis are at best defined arbitrarily, or at worst defined based on observed distributions of cancer outcomes, leading to the silent multiple comparisons problem described above. By identifying putative source(s) of hazard a priori, we can pose and investigate questions regarding the causal effects of these exposures on cancer incidence, and the statistical hypotheses can be formed around the exposed population and time period, thus avoiding the pitfalls of the current approach. 


We propose the use of a causal inference approach rooted in the potential outcomes framework \citep{rubin1974estimating} to evaluate the strength of evidence that a given exposure caused increased risk of certain cancers in the exposed population and time period. We define a causal estimand for this context called the causal SIR (cSIR). The cSIR is the following ratio: the expected cancer incidence in the exposed population divided by the expected cancer incidence in the same population under the counterfactual scenario of no exposure. We also lay out the identifying assumptions required to estimate the cSIR from observed data. Estimating causal effects with observational data requires rigorous approaches for eliminating confounding bias. Literature in the field of causal inference now provides methods that can manipulate observational data to remove confounding (on observed confounders), thereby mimicking a randomized experiment, so that the causal effect of exposure may be extracted \citep{rubin1973matching,rosenbaum1983central,rosenbaum1984reducing,hernan2006estimating,stuart2010matching}.


\subsection{Case study and approach to cSIR estimation}\label{sss:data}

To clarify our approach to cSIR estimation, we now introduce the cancer cluster investigation that will be used as a case study in this paper. However, an analogous approach to estimation can be used in nearly any cancer cluster investigation. Endicott, New York (NY) was the home of the first IBM manufacturing complex. A spill of thousands of gallons of a mixture of chemicals by IBM in 1979 has plagued the town for decades. According to the NY State Department of Environmental Conservation (DEC) \citep{dec2002village}, trichloroethylene (TCE), a metal degreaser and a known carcinogen, was the spilled contaminant that migrated the furthest outside the IBM plant and into the surrounding community, carried via groundwater. In 2002, an investigation mandated by the DEC discovered that the TCE that had migrated into the soil in residential areas was evaporating and the resultant vapors entering indoor air in homes at dangerous levels, a process known as vapor intrusion. How long prior to 2002 the community had been exposed to TCE vapor intrusion remains unknown. TCE exposure is known to cause kidney cancer, but evidence in human studies has also suggested associations with lymphomas and childhood leukemia and liver, biliary tract, bladder, esophageal, prostate, cervical, and breast cancers \citep{epa2011tce}.

In 2006, the NY State Department of Health (DOH) conducted an investigation of cancer rates in Endicott between 1980-2000 using a SIR analysis. They found rates of kidney and testicular cancer were significantly higher than background rates \citep{doh2006health}. To our knowledge, no follow up investigation has been conducted to determine whether residential exposure to the TCE vapors, detected in 2002, after the end date of the DOH study, has led to increased cancer rates in the community. We wish to estimate the cSIR for kidney/renal pelvis cancer and for bladder cancer in the TCE-exposed population of Endicott in order to investigate whether TCE exposure caused increased incidence of these cancers in Endicott during 2005-2009.


To estimate the cSIRs, we need information about what the incidence of these cancers in the TCE-exposed portion of Endicott would have been both with and without the TCE exposure. Of course, the observed cancer incidence in Endicott provides information about the former, since exposure to TCE was the factual, observed scenario. In order to learn about what the cancer incidence might have been in the counterfactual scenario of no exposure, we must obtain cancer data from unexposed populations that are as similar as possible to the exposed population in Endicott in terms of potential confounders of the causal effect of TCE exposure on cancer outcomes. To identify these populations, we will rely on causal matching.

Matching is one of the most well established causal inference approaches for eliminating confounding in observational data \citep{rubin1973matching,rubin2000combining,abadie2006large,ho2007matching,stuart2010matching,abadie2011bias,iacus2011multivariate}. Matching methods are nonparametric and require minimal assumptions. For each exposed unit in the data, matching methods identify a fixed number of ``matched control'' units that are not exposed to the hazard but are as similar as possible to the exposed unit in terms of observed confounders. Typically only the matched units are retained for analysis, unmatched units are discarded. Under certain assumptions, this procedure ensures that the distributions of confounders are similar in the exposed and unexposed groups, as in a randomized experiment. Matching is known as a ``design phase method'', because it aims to remove confounding effects without invoking outcome data. Procedures applied subsequently using the outcomes to estimate causal effects are known as ``analysis phase methods''. The idea to apply matching in SIR analyses was introduced by \cite{dominici2007role}.

\begin{figure}[h]
\centering
\includegraphics[scale=.6,page=3]{endicott_maps.pdf}
\includegraphics[scale=.6,page=1]{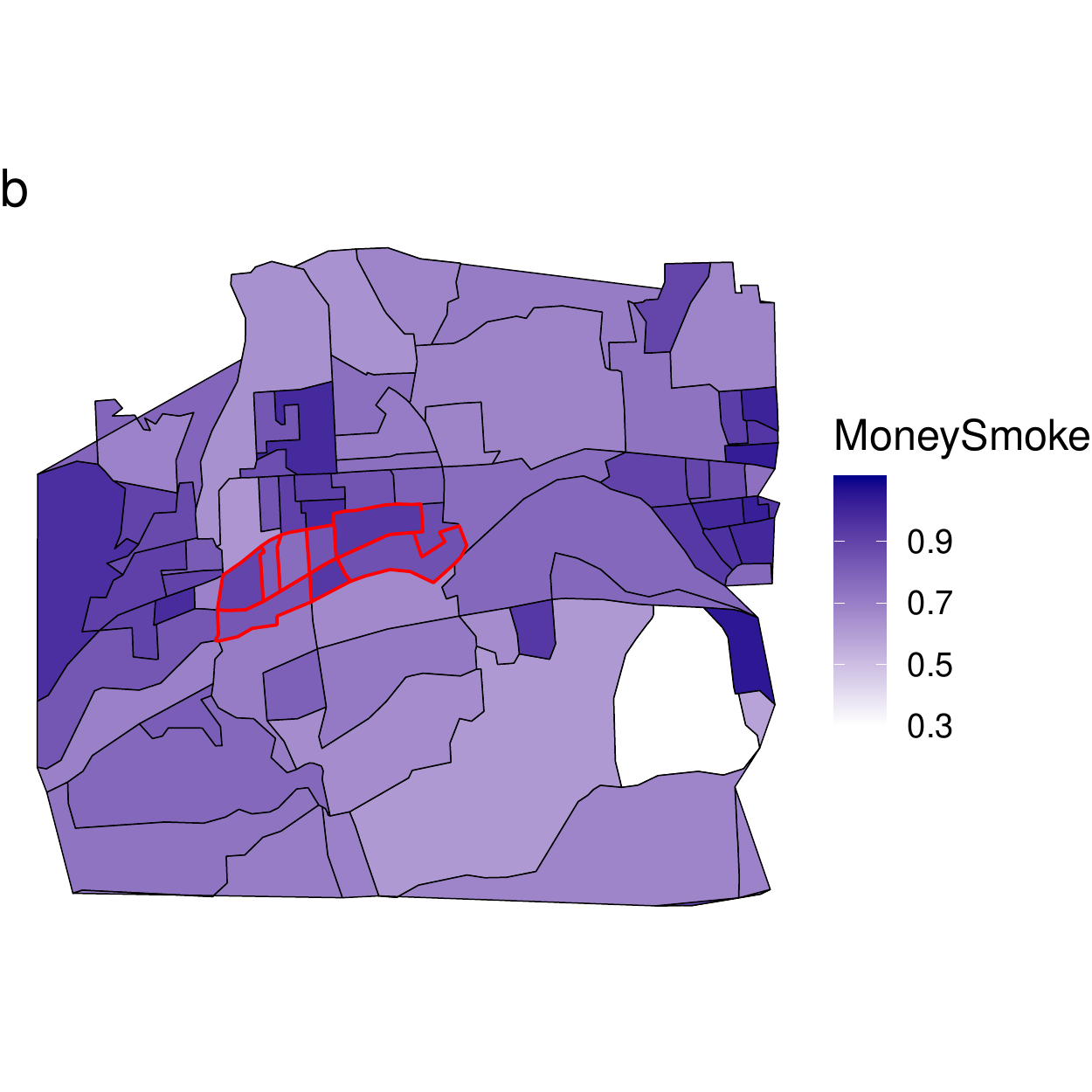}
\caption{Bladder cancer incidence rate per 100,000 population 2005-2009 by census block group for the TCE-exposed area of Endicott (red borders) and surrounding area (a) and money spent on smoking products as a portion of per capita income (b).  
}
\label{fig:maps}
\end{figure}

We partition the exposed area (Endicott) into exposed sub-units (census block groups; CBGs), and these will be our units of analysis.  Furthermore, for each of the exposed sub-units we assume that both cancer incidences are known (in many cases, the exact locations of diagnoses will be reported by the concerned community) and that confounder data is publicly available. Figure ~\ref{fig:maps} provides maps of the bladder cancer incidence and the money spent on smoking products (a potential confounder) by CBGs both within the Endicott area (red lines) and for the CBGs in the surrounding area. The similar spatial patterns in the cancer incidence and confounder maps are noticeable. Our goal is to identify matched controls for the sub-units within the exposed area. 

We consider this spatial partitioning for two important reasons. First, matching on potential confounders for the sub-units (e.g. CBGs) is more likely to eliminate confounding bias than matching at the higher level of spatial aggregation (e.g. the whole exposed area), as many of these potential confounders may vary widely within the exposed area. Second, this partitioning provides a larger sample of exposed sub-units within the exposed area, thus allowing for the application of classic statistical models to the data. In general, it is advisable to partition the exposed area into small administrative units for which confounder data are readily available, e.g., zip codes, census tracts, or CBGs.

Nation-wide confounder data (e.g., socioeconomic and demographic variables) at the CBG level are publicly released by the US Census Bureau. For many types of environmental hazard, relevant geocoded contaminant and toxic chemical use information is available from the Environmental Protection Agency (EPA) (see Section 4 for more information). Therefore we assume that exposure and confounder data are both available for all CBGs and similar small administrative units nation-wide. Obtaining small area cancer incidence data, on the other hand, is challenging. 

Cancer registries in most states collect detailed data on nearly all patients at the time of diagnosis, including age, sex, address, and diagnosis codes. While improvements are being made in the accessibility of these data, privacy concerns mean that these data, even when aggregated over small geographic areas, may not be available to researchers without access to high security servers. Moreover, the process of requesting and obtaining access to the data can take many months, delaying time-sensitive work. The high degree of aggregation in most publicly available cancer incidence data often renders it too coarse to perform the desired analyses. In the past, some cancer cluster investigations have relied on publicly available data, while others have endured the process of obtaining the more detailed data from registries. To make our methods as accessible as possible, they will rely solely on publicly available cancer incidence data.

The most commonly used publicly available cancer incidence data are collected and published by the Surveillance, Epidemiology, and End Results (SEER) Program of the US National Cancer Institute \citep{seer2018}. Today, SEER compiles cancer diagnosis records from a total of 18 state and city cancer registries. Each of these registries has contributed data dating back to 2000 or prior. For each record in the SEER data, demographic data and diagnosis information are provided, as well as the year of diagnosis and the person's county of residence at the time of diagnosis. Thus, the lowest available level of spatial aggregation for these data is the county level.

To our knowledge, only two states have made small area cancer data publicly available. Illinois provides state-wide zip code level cancer incidence data for 11 anatomic site groupings \citep{ilcancer}; however kidney and bladder cancer incidences are combined with all other unrinary tract cancers. NY provides state-wide CBG level cancer incidences for 23 different anatomic sites, including kidney and bladder (separately), over the period 2005-2009 \citep{boscoe2016public}. To ensure the protection of privacy, these publicly available small area cancer incidence data are aggregated over fairly large time periods, multiple disease classifications, and demographic strata, which may render the incidence values not directly comparable to those of interest in the exposed community. We refer to the set of regions from which we have either SEER data or relevant small area cancer incidence data as SEER+. For our Endicott case study, SEER+ includes all SEER-participating regions and all of NY state.

Thus, after partitioning the exposed area of Endicott into CBGs, we seek matched control CBGs within SEER+. In the matched dataset, we are faced with the challenge of having cancer incidence data for (some or all of) the matched controls at the county level while our analysis is carried out at the smaller CBG level. This problem is referred to as spatial over-aggregation. We have developed a novel Bayesian model to be applied to the matched data that resolves the spatial over-aggregation and estimates the cSIR. This model jointly (1) imputes the cancer incidence in matched control CBGs by borrowing information from the limited publicly available small area cancer incidence data, while imposing the constraint that CBG incidences should sum to their encompassing county's observed incidence and (2) fits a log-linear model to the cancer incidences in the matched data, combining the imputed incidences for the matched controls and observed incidences in the community under study to estimate the cSIR. By fitting the model jointly, we are able to account for the additional uncertainty imposed on the cSIR estimate from the use of predicted CBG cancer incidences.

In Section 2, we formally define the cSIR and lay out our proposed estimation procedure. In Section 3, we use simulations to compare our methods to the existing SIR analysis methods used in cancer cluster investigations. In Section 4, we describe the analysis and results of the Endicott case study. Finally, we discuss our findings and conclude in Section 5.

\section{Methods}\label{s:methods}

\subsection{The potential outcomes framework and notation}

The population of interest in this context is the population exposed to the source of hazard \textit{within the concerned community under scrutiny}. It is important to note that we wish to estimate the effect of exposure only in this single exposed population. Without loss of generality, we assume that, as in our case study, the exposed region is partitioned into its component CBGs. In defining the cSIR and its identifying assumptions and formalizing the statistical models, CBGs are the units of analysis.

We now define notation that will be used to develop the estimand and methods. Causal inference methods are often situated within the potential outcomes framework, as defined by \cite{rubin1974estimating}, which we will now adapt to the cancer cluster analysis setting. Throughout this section, we utilize three different subscripts to index different datasets. Subscript $h=1,...,H$ indexes the CBGs of the exposed region, subscript $i=1,...,N$ indexes the matched dataset (i.e., the exposed region CBGs and the matched control CBGs), and $j=1,...,J$ indexes the data used to build the model for prediction of the cancer incidence in the matched controls. Let $Y_h$ be a random variable representing the observed cancer incidence in CBG $h$ during the time period of interest, $h=1,...,H$. We let $T_h$ denote an indicator of exposure status, with $T_h=1$ for all $h=1,...,H$ because each CBG in this set was exposed to source of hazard. Let $\boldsymbol{X}_h$ be a vector of observed confounder values for CBG $h$. Then the potential outcomes are $Y_h(T=1)$, the cancer incidence that would have been observed in CBG $h$ under exposure to the source of hazard, and $Y_h(T=0)$, the cancer incidence that would have been observed in CBG $h$ under no exposure.

The fundamental problem of causal inference is that, at most, only one of the two potential outcomes can ever be observed for a given unit, either its outcome under exposure or its outcome under no exposure. In this case, within the population of interest, we only observe outcomes under exposure to the source of hazard. The unobserved potential outcome is called the counterfactual. As in nearly all causal inference analyses, we invoke the stable unit treatment value assumption (SUTVA) in order to ensure the existence of the potential outcomes \citep{rubin1980randomization}. SUTVA requires that the exposure be well-defined, i.e. that there is only a single ``version'' of exposure, and that the exposure status of a given unit does not affect the outcome of other units. See the Discussion section for an evaluation of the appropriateness of the SUTVA assumptions in the cancer cluster setting and ideas for how to extend our methodology to account for violations of these assumptions.

\subsection{The causal SIR and identifying assumptions}\label{ss:csirassumptions}

Using the potential outcomes, we now define the cSIR as:
\[ cSIR=\frac{E\left[Y(T=1) | T=1\right]}{E\left[Y(T=0) | T=1 \right]} \]
Again, we only wish to estimate the cSIR for the exposed population in the community under study. The cSIR is a ratio analogue to the average treatment effect on the treated, which is a commonly used causal inference estimand. 
As with the classic SIR analysis, we are interested in evaluating the strength of evidence that $cSIR \neq 1$, with $cSIR=1$ equivalent to $E\left[Y(T=1) | T=1\right]=E\left[Y(T=0) | T=1\right]$, i.e., no causal effect of exposure in the exposed population.

Now, drawing on additional data from regions unexposed to the same type of hazard as the population under study, the cSIR can be estimated under the following identification assumptions. These assumptions are nearly identical to those needed to estimate the average treatment effect on the treated-- ignorability and causal consistency. First, cSIR identification relies on the assumption of no unobserved confounding, stated mathematically as $T \independent Y(T=0)|\boldsymbol{X}$, i.e., conditional on observed confounders, $\boldsymbol{X}$, the exposure assignment is independent of the potential outcome under no exposure. The assumption of no unobserved confounding is untestable, and its plausibility must be assessed based on subject matter knowledge. Moreover, we must assume positivity, stated mathematically as $P(T=1|\boldsymbol{X})<1$. The assumption of positivity requires that every CBG in the exposed area could feasibly have been unexposed. The plausibility of the positivity assumption can often be evaluated by determining whether suitable unexposed matches can be found for each exposed unit. If unexposed units exist that are highly similar to an exposed unit in terms of all other relevant features, then this provides evidence that the exposed unit could feasibly have been unexposed. Together, the assumptions of no unobserved confounding and positivity are known as ignorability. Finally, the causal consistency assumption states that $Y=Y(T=1)\times T-Y(T=0)\times(1-T)$, i.e. the observed outcome is equal to the potential outcome under the observed exposure level.

By applying these assumptions, we can see that the cSIR is identifiable from the observed data. Consider the numerator of the SIR, $E\left[Y(T=1) | T=1\right]$. Note that
\[E\left[Y(T=1) | T=1\right]=E_X\left[E\left[Y(T=1) | T=1, X\right]\right]=
E_X\left[E\left[Y | T=1, X\right]\right]\]
where the last equality holds by causal consistency. Similarly, for the denominator, $E\left[Y(T=0) | T=1\right]=E_X\left[E\left[Y(T=0) | T=1, X\right]\right]$. Now, we invoke the ignorability assumption, which states that treatment status is independent of the potential outcomes conditional on confounders, so that\\$E\left[Y(T=0) | T=1, X\right]=E\left[Y(T=0) | T=0, X\right]$. Thus, we have
\[E\left[Y(T=0) | T=1\right]=E_X\left[E\left[Y(T=0) | T=0, X\right]\right]=E_X\left[E\left[Y | T=0, X\right]\right]\]
by applying causal consistency as above. Thus, we see that both the numerator and denominator of the cSIR are identifiable and can be estimated with observed data.

\subsection{Estimation of the cSIR: design phase}\label{csir_est:design}
To estimate the cSIR, matching is used in the design phase to remove confounding, followed by a Bayesian estimation procedure in the analysis stage to appropriately account for all sources of uncertainty. We begin by describing the design phase of our approach below.

The goal of our matching procedure is to obtain a set of unexposed CBGs with distributions of the potential confounders as similar as possible to the distributions of the potential confounders in the CBGs within the exposed area. Thus, the first step in the design phase is to identify a (hopefully large) set of unexposed CBGs in which to search for matches. Certain contaminants (such as air pollutants) may be assumed to be universally present at some low level, making it impossible to find truly unexposed CBGs. In such cases, the ``controls'' selected for matching could instead be CBGs with exposure levels below a certain threshold, where the threshold should be selected so that exposures below that level are believed to have no effect on health.

 We recommend utilizing the matching procedure that provides the best covariate balance between the exposed and unexposed regions, i.e. the smallest standardized differences in means. Covariate balance is an indication that, as in a randomized trial, the distributions of observed confounders are similar in the exposed and unexposed groups. Many matching procedures allow for ratio matching-- finding multiple matches for each exposed region, and, because small area cancer incidence rates are often unstable, we suggest applying ratio matching in this context in order to obtain as much information as possible about the expected cancer incidence under no exposure.

\subsection{Estimation of the cSIR: analysis phase}\label{csir_est:analysis}

After matching, if the cancer incidence data for both exposed and matched control CBGs were available at the CBG level, a loglinear modeling approach could be used to estimate the cSIR. For clarity, we first describe the model that would be applied in this setting. The loglinear model approach for estimating disease risk relative to a point source with aggregated data was introduced by \cite{diggle1997regression} from a frequentist perspective and by \cite{wakefield2001bayesian} from a Bayesian perspective. In this context, the model should be fit using data from both the exposed CBGs and the matched control CBGs and should include both exposure status and confounder variables as predictors. If matching procedures are entirely successful at removing all differences in confounder distributions between the exposed and unexposed, then adjustment for the confounders in the analysis phase is not needed. However, in practice, matching typically does not remove these differences entirely, thus adjustment for the confounders in analysis phase modeling is recommended.

Let $i$ ($i=1,...,N$) index the matched dataset. The loglinear model has the form
\[
\text{log}(E\left[Y_i\right])=\alpha_0+\alpha_1 T_i + \boldsymbol{\alpha}_2\boldsymbol{X}_i+\text{log}(P_i)
\]
where $P_i$ is the population size in CBG $i$, and $\text{log}(P_i)$ is an offset term used to account for potential differences in population size across the CBGs. If the cancer incidence data are collected over different time periods for some of the matched control CBGs, the offset could represent person-time rather than population size. Because the sample size for this model, $N$, will generally be small, a Bayesian approach to model fitting will generally provide more stable estimates than frequentist models. The cSIR estimate is $\text{exp}(\hat{\alpha}_1)$, and uncertainties and confidence regions follow accordingly. If cancer incidence data exhibit excess zeros or extra-Poisson variation, existing extensions to the Bayesian Poisson regression model to accommodate these deviations could be used.

We now introduce our approach to estimate the cSIR when the cancer incidence data for some or all of the matched controls CBGs are over-aggregated to the county level, as described in Section~\ref{sss:data}. In this setting, we propose a two-stage Bayesian model to be applied to the matched dataset. This model (1) combines publicly available CBG cancer data, SEER data, and socio-economic and demographic data to predict cancer incidence in the matched control CBGs and (2) using these predictions and the observed incidence data from the exposed CBGs, fits the loglinear model described above to estimate the cSIR.

\subsubsection{Stage 1: prediction model}
The goal of the prediction stage is to use the publicly available NY CBG cancer incidence data to model relationships between CBG incidence and community characteristics and to apply that model to predict cancer incidences in the matched control CBGs, taking into account the additional information provided by the observed SEER county level cancer incidences. In order to account for the observed county level cancer incidences, our model must incorporate the constraint that the CBG predicted incidences within a given county should sum to the observed county incidence. Finally, because this model is fit only to NY data but is employed for prediction of CBG cancer incidences in other states, we must make an additional assumption that the results of this model are transportable, or equivalently, that the model has external validity \citep{singleton2014motivating}.

Subscript $j=1,...,J$ is used to index the set of all NY CBGs. Let $\boldsymbol{Z}_j$ denote the vector of predictors to be included in the prediction stage for CBG $j$. The variables in $\boldsymbol{Z}$ should include all the confounders in $\boldsymbol{X}$, but may include additional variables that are predictive of cancer incidence but are not believed to be confounders. We assume $Y_j \sim \text{Poisson}(\lambda_{j})$ and $\text{log}(\lambda_{j})=\boldsymbol{Z}_{j}\boldsymbol{\beta}+\text{log}(P_j)$, where $P_j$ is the population size (or person-time, if needed). We denote by $\psi (j)$ the set of indices for all CBGs in the same county as CBG $j$, and $\boldsymbol{Y}_{\psi (j)}$ the vector of all CBGs in the same county as CBG $j$. Then, it is a well-known result that $\left(\boldsymbol{Y}_{\psi (j)}|\sum_{l \in \psi (j)} Y_l=K \right) \sim \text{Multinomial}(K,\boldsymbol{\pi})$,
with $K$ the observed cancer incidence in the county containing CBG $j$ and $\boldsymbol{\pi}$ a vector whose elements are the proportions of the $K$ cancer cases that fall into each of the county's CBGs. Thus, in order to have our model account for the constraint that CBG incidence predictions should sum to their county's observed incidence, we will develop the model around a multinomial likelihood. For each $Y_j$ we have corresponding multinomial distribution parameters $K_j$ ($K_j$ is the same for all CBGs from the same county) and $\pi_j$, where $\pi_j$ is the proportion of the cancer incidence in its encompassing county that falls into CBG $j$. Note that, by the same distributional result given above, 
\[\pi_j=\frac{\lambda_j}{\sum_{l \in \psi (j)} \lambda_l}\]
and this property dictates the form of the prediction model.

The prediction model is similar to a classic loglinear model, but includes a non-traditional offset that imposes the constraint that the estimated multinomial probabilities must sum to one. It follows from the properties laid out above that the $\pi_{j}$ should have the following relationship to the predictors:
\[
\text{log}(\pi_{j})=\boldsymbol{Z}_{j}\boldsymbol{\beta}+\text{log}(P_j)-\text{log}(\sum_{l \in \psi (j)} e^{\boldsymbol{Z}_{l}\boldsymbol{\beta}}P_l)
\]
where the final term is an offset which imposes the constraint. Note that this implies that
\[
\pi_{j}=\frac{e^{\boldsymbol{Z}_{j}\boldsymbol{\beta}}P_j}{\sum_{l \in \psi (j)} e^{\boldsymbol{Z}_{l}\boldsymbol{\beta}}P_l}
\]
so that the CBG probabilities within a county sum to one, as desired. This results in the following data likelihood:
\[
L(\boldsymbol{\beta}|\boldsymbol{Y},\boldsymbol{Z})=
\prod_{j=1}^{J} (K_j!)^{1/||\psi (j)||} \frac{ (e^{\boldsymbol{Z}_{j}\boldsymbol{\beta}}P_j)^{Y_{j}}}{Y_{j}!(\sum_{l \in \psi (j)} e^{\boldsymbol{Z}_{l}\boldsymbol{\beta}}P_l)^{Y_{j}}}
\]
where $||\psi (j)||$ denotes the cardinality of $\psi (j)$. Using a Bayesian approach, we can fit this model to the NY CBG cancer incidence data through the use of a simple Metropolis sampler. The resulting posterior summaries of $\boldsymbol{\beta}$ speak to the associations between a CBG's features and the proportion of the cancer incidence of its larger county that it accounts for.

Moreover, for any CBG in the SEER states, where $\boldsymbol{Z}_{j}$ and $K_j$ are known, we can obtain posterior predictive samples of its cancer incidence from the corresponding multinomial distribution. We note that we only need posterior predictive samples from the matched control CBGs in order to do cSIR estimation. However, because the model relies on normalization of the CBG proportions within counties, in order to obtain the multinomial posterior predictive samples for any CBG we must utilize the predictor data from all the other CBGs in its encompassing county, as well as the observed SEER incidence for the county. For a given matched control CBG, we denote the posterior predictive samples of its cancer incidence as $\left\lbrace Y^{(1)},...,Y^{(M)}\right\rbrace$, and these get passed into the estimation stage of the model. 

\subsubsection{Stage 2: estimation model}
Using the observed incidences in the exposed CBGs and the posterior predictive samples of the incidence in the matched control CBGs, we estimate the cSIR in the second stage of the model. As above, we employ a Bayesian loglinear model, now integrating in the sampled outcomes for the controls at each iteration of the sampler. By including the full distribution of predicted cancer incidences in the estimation stage, rather than a single summarized predicted value, our cSIR estimate will capture the additional variability generated by the use of predicted cancer incidences for the matched controls.

Now utilizing only the matched data, let 
  \[
    \tilde{Y}_i^{(m)} = \left\{\begin{array}{lr}
        Y_i, & \text{if } T_i=1\\
        Y_i^{(m)}, & \text{if } T_i=0\\
        \end{array}\right.
  \]
 $i=1,...,N$. Then in each iteration of the Metropolis sampler for the estimation model, we plug in a different $\tilde{Y}_i^{(m)}$ sample, i.e. for $m=1,...M$ we collect a posterior sample of $\left\lbrace \alpha_0, \alpha_1, \boldsymbol{\alpha}_2 \right\rbrace$ from
\[
\text{log}(E\left[\tilde{Y}_i^{(m)}\right])=\alpha_0+\alpha_1 T_i + \boldsymbol{\alpha}_2\boldsymbol{X}_i+\text{log}(P_i).
\]
The cSIR and its uncertainties are estimated from this model as described above.

We note that, in its current form, stage 1 of this method relies on the Poisson distribution and is not equipped to handle zero-inflated cancer data. Thus, the reasonableness of a Poisson data likelihood should be assessed before applying this method to data. Moreover, prediction models based on publicly available data have not yet been validated for any cancer type and the results and model fit should be evaluated on a case-by-case basis. Validation of these prediction models is an important topic for future work.

\section{Simulations}
In this section, simulations are conducted to compare the CDC's SIR analysis to our proposed method. Our intent is to demonstrate how the use of matched controls and Bayesian estimation methods, under the assumptions laid out above, leads to stable and unbiased estimation of the effect of an exposure on cancer incidence. All simulations are carried out in R statistical software \citep{r2018}, and code is available at \url{https://github.com/rachelnethery/causalSIR}.

\subsection{Simulation Structure}

The simulations are constructed using real confounder data from the SEER states so that we are ensured that the simulated data reflect the complexity of real data. For each CBG within each region covered by SEER, we obtain data on the following variables (in parentheses, the names used for the remainder of the paper): percent of the population age 65+ (P65+), percent of the population male (PMale), percent of the population white (PWhite), percent of the adult population unemployed (Unemploy), average commute time (Commute), median household income (Income), dollars spent on smoking products as a portion of per capita income (MoneySmoke), and percent of total dollars spent on food that was spent on food outside the home (MoneyFood). All variables come from ESRI Business Analyst \citep{esri2018}. 

 Let $\boldsymbol{X}$ denote a matrix containing this set of confounders for each CBG in the SEER states, $\boldsymbol{P}$ denote a vector of the population in each CBG, $\boldsymbol{T}$ denote the vector of exposure indicators for each CBG, and $\boldsymbol{Y}$ denote the vector of cancer incidences for each CBG. We generate $\boldsymbol{T}$ and $\boldsymbol{Y}$ from the models \[\text{logit}(P(\boldsymbol{T}=1))=\gamma_0+\boldsymbol{\gamma}_1 \boldsymbol{X}\]
 \[\text{log}(E\left[\boldsymbol{Y}\right])=\alpha_0+\alpha_1 \boldsymbol{T} + \boldsymbol{\alpha}_2\boldsymbol{X}+\text{log}(\boldsymbol{P})\]
using different specifications of the parameter values to obtain different simulated conditions. Through these specifications, we produce the following four simulation structures:
\begin{enumerate}
    \item no exposure effect ($\boldsymbol{T}\nrightarrow \boldsymbol{Y}$) and no confounding ($\boldsymbol{X}\nrightarrow \boldsymbol{T},\boldsymbol{X}\nrightarrow \boldsymbol{Y}$)
    \item no exposure effect ($\boldsymbol{T}\nrightarrow \boldsymbol{Y}$) and confounding ($\boldsymbol{X}\rightarrow \boldsymbol{T},\boldsymbol{X}\rightarrow \boldsymbol{Y}$)
    \item exposure effect ($\boldsymbol{T}\rightarrow \boldsymbol{Y}$) and no confounding ($\boldsymbol{X}\nrightarrow \boldsymbol{T},\boldsymbol{X}\rightarrow \boldsymbol{Y}$)
    \item exposure effect ($\boldsymbol{T}\rightarrow \boldsymbol{Y}$) and confounding ($\boldsymbol{X}\rightarrow \boldsymbol{T},\boldsymbol{X}\rightarrow \boldsymbol{Y}$)
\end{enumerate}
Moreover, within simulations 3 and 4, we simulate data exhibiting varying strengths of exposure effects to test the power of our method to detect effects, an important consideration given that the number of CBGs containing the exposed population is generally small (i.e., 10 or fewer). See Table~\ref{sim_struct} for the parameter values used to construct each simulation. Within each of the simulation scenarios, we generate 5,000 datasets, with $\boldsymbol{T}\sim \text{Bernoulli}(P(\boldsymbol{T}=1|\boldsymbol{X}))$ held fixed across simulations but a different $\boldsymbol{Y}\sim \text{Poisson}(E\left[ \boldsymbol{Y} | \boldsymbol{X}, \boldsymbol{T} \right])$ simulated in each. 10 exposed CBGs ($T=1$) are randomly selected to represent the exposed population of interest, and these are also fixed across simulations.

\begin{table}[h!]
\centering
\caption{Parameter values used to simulate data. The values are chosen to reflect the anticipated magnitude and direction of the relationship between each variable and the exposure/outcome.}
\begin{tabular}{llrrR{3cm}R{3cm}}
  \hline
 & & Sim 1 & Sim 2 & Sim 3 & Sim 4 \\ 
  \hline
 \multirow{9}{*}{Exposure Model} & Intercept ($\gamma_0$) & -1.15 & 0 & -1.15 & 0 \\ 
 & MoneyFood  ($\gamma_{11}$)  & 0 & 0.0009    & 0 & 0.0009 \\ 
 & MoneySmoke  ($\gamma_{12}$) & 0 & 0.015     & 0 & 0.015 \\ 
 & P65+        ($\gamma_{13}$) & 0 & 0.003     & 0 & 0.003 \\ 
 & PMale      ($\gamma_{14}$)  & 0 & -0.001    & 0 & -0.001 \\ 
 & PWhite     ($\gamma_{15}$)  & 0 & -0.01     & 0 & -0.01 \\ 
 & Unemploy   ($\gamma_{16}$)  & 0 & 0.004     & 0 & 0.004 \\ 
 & Commute     ($\gamma_{17}$) & 0 & 0.002     & 0 & 0.002 \\ 
 & Income     ($\gamma_{18}$)  & 0 & -0.01     & 0 & -0.01 \\ 
 \hline
  \multirow{10}{*}{Outcome Model} & Intercept ($\alpha_0$) & -5.99 & -5 & -5 & -5 \\ 
 & Exposure   ($\alpha_1$)  & 0 & 0     & $\left\{\text{log}(1.1),\text{log}(1.2),\right.$ $\left. ...,\text{log}(2)\right\}$ & $\left\{\text{log}(1.1),\text{log}(1.2),\right.$ $\left. ...,\text{log}(2)\right\}$ \\ 
 & MoneyFood   ($\alpha_{21}$) & 0 & 0.007 & 0.007     & 0.007 \\ 
 & MoneySmoke  ($\alpha_{22}$) & 0 & 0.015 & 0.015     & 0.015 \\ 
 & P65+         ($\alpha_{23}$)& 0 & 0.03  & 0.03      & 0.03 \\ 
 & PMale      ($\alpha_{24}$)  & 0 & -0.001& -0.001    & -0.001 \\ 
 & PWhite      ($\alpha_{25}$) & 0 & -0.02 & -0.02     & -0.02 \\ 
 & Unemploy    ($\alpha_{26}$) & 0 & 0.004 & 0.004     &  0.004\\ 
 & Commute     ($\alpha_{27}$) & 0 & 0.002 & 0.002     &  0.002\\ 
 & Income      ($\alpha_{28}$) & 0 & -0.005& -0.005    &  -0.005\\ 
 \hline
  \end{tabular}
\label{sim_struct}
\end{table}

\subsection{Methods Compared}\label{ss:simmethods}

Once we have simulated exposure and outcome data for each CBG in the SEER states, we analyze the data using three different methods: CDC's recommended SIR analysis (abbreviated as CDC), a variant of the CDC's method that employs Poisson regression modeling (abbreviated as PR), and our proposed cSIR analysis (abbreviated as cSIR). We formalize the CDC's SIR analysis here. Let $D$ represent the observed cancer incidence in the concerned community during the relevant time period. An expected number of cancer cases $E=E(D)$ is computed for the community based on the incidence in a background population. Then, the SIR is $S=\frac{Y}{E}$. Using the assumption that $D \sim Poisson(S\times E)$, confidence intervals are computed by invoking the relationship between the Poisson and Chi-Square distributions \citep{sahai1993confidence}.

Note that our proposed method and the CDC's protocol for SIR estimation differ in their approach to identifying the population and time period under study. Our proposed approach makes these determinations by identifying the population and time period exposed to a pre-specified source of hazard. The CDC protocol bases their considerations only on where/when the high cancer incidence is reported. In practice, these procedures would likely lead to a different population/time period under study but in our simulations, we use the same population/time period under study in all methods for comparability (i.e., the 10 exposed CBGs selected as described above).

The CDC method is implemented by using the simulated cancer incidences from all the SEER state CBGs outside the community of interest to compute the expected incidence (i.e., background rate). The PR variant of the CDC's method does make some effort at adjustment for confounding-- a frequentist Poisson regression model is fitted to the data from all the CBGs, using the confounding variables as covariates, but estimating the SIR as the exponentiated parameter estimate corresponding to an indicator of inclusion in the community of interest, rather than the true exposure indicator. Finally, we implement cSIR, assuming appropriate spatial aggregation of the cancer incidence data. We identify matched controls for each exposed CBG using 20:1 mahalanobis distance nearest neighbor ratio matching and then fit the Bayesian loglinear model.

\subsection{Simulation Results}

\begin{figure}[h!]
\centering
\includegraphics{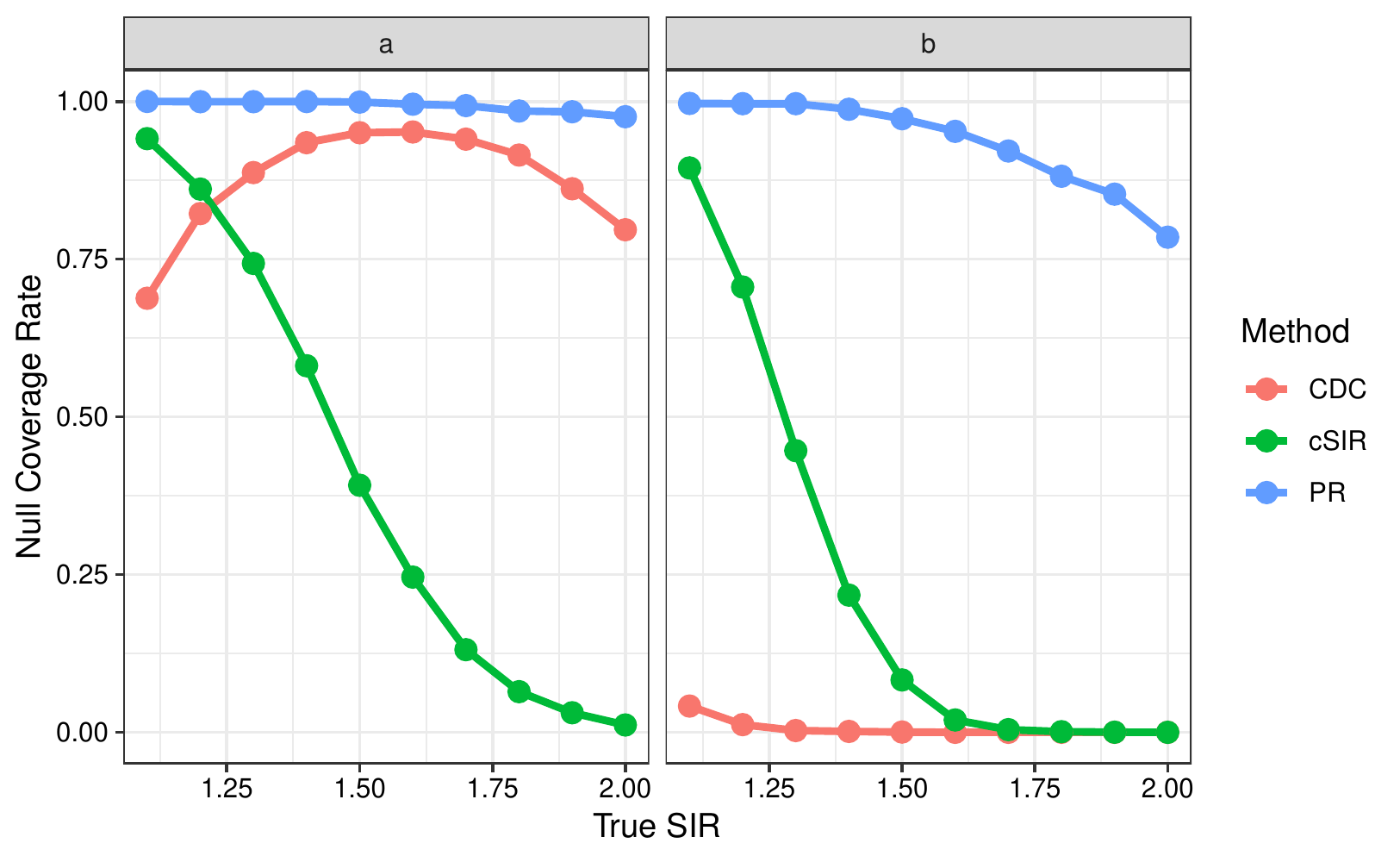}
\caption{Trends in the rate of coverage of the null SIR as the true SIR increases in simulation 3 (a) and simulation 4 (b). This reflects the power of each method to detect exposure effects.}
\label{fig:null_covg}
\end{figure}

Results are given in Table~\ref{tab:sim_results}, which provides for each method the bias in the point estimate, the coverage rate of the true SIR for 95\% confidence/credible intervals, and the width of the 95\% confidence/credible intervals. Note that, because bias and coverage results are consistent across different exposure effect values ($\alpha_1$) in simulations 3 and 4, only select results are shown. Figure~\ref{fig:null_covg} shows the rate of coverage of the null SIR value, 1, for each method as the true SIR value increases from 1.1 to 2 in simulations 3 and 4. These results reflect the power of each method to detect non-null exposure effects. (In simulations 1 and 2 where there is no exposure effect, coverage of the null is the same as coverage of the true SIR.)

\begin{table}[h!]
\centering
\caption{Simulation results comparing the proposed cSIR method with the standard cancer cluster SIR estimation method (CDC) and a similar Poisson regression approach (PR). Shown are the bias in the point estimate, the coverage rate of the true SIR for 95\% confidence/credible intervals, and the width of the 95\% confidence/credible intervals.}
\begin{tabular}{rrrp{1.5cm}p{2cm}p{2cm}}
  \hline
 & True SIR & Method & Bias & Coverage True SIR & CI Width \\ 
  \hline
 \multirow{3}{2.5cm}{Simulation 1} & \multirow{3}{*}{1} & CDC & -0.00 & 0.96 & 0.70\\ 
 &  & PR & 0.20 & 0.99 & 4.81 \\
 &  & cSIR & -0.02 & 0.94 &  0.66 \\  \hline
  \multirow{3}{2.5cm}{Simulation 2} & \multirow{3}{*}{1} & CDC & 0.55 & 0.13 & 0.79 \\ 
 &  & PR & -0.27 & 0.99 & 5.86  \\ 
 &  & cSIR & -0.00 & 0.95 & 0.50 \\ \hline
  \multirow{9}{2.5cm}{Simulation 3} & \multirow{3}{*}{1.1}& CDC & -0.33 & 0.43 & 0.54 \\ 
  & & PR & -0.06 & 1.00 &  12.89 \\ 
  &  & cSIR & -0.03 & 0.95 &  0.73\\ \cline{2-6}
   & \multirow{3}{*}{1.5} & CDC & -0.54 & 0.09 & 0.58 \\ 
 & & PR & 1.23 & 1.00 & 49.08 \\ 
  &  & cSIR & -0.05 & 0.94 & 0.87 \\ \cline{2-6}
   & \multirow{3}{*}{2} & CDC & -0.84 & 0.00 & 0.60 \\ 
  & & PR & 0.73 & 1.00 & 49.08 \\ 
   & & cSIR & -0.07 & 0.93 & 1.02 \\ \hline
   \multirow{9}{2.5cm}{Simulation 4} & \multirow{3}{*}{1.1} & CDC & 0.56 & 0.13 & 0.80 \\ 
 & & PR & 0.02 & 1.00 & 14.29 \\ 
  &  & CSIR & -0.01 & 0.95 & 0.53 \\ \cline{2-6}
 & \multirow{3}{*}{1.5} & CDC & 0.51 & 0.26 & 0.83 \\ 
 & & PR & 3.76 & 0.99 & 112.96 \\ 
 & & CSIR & -0.01 & 0.94 & 0.62 \\ \cline{2-6}
& \multirow{3}{*}{2} & CDC & 0.35 & 0.58 & 0.84 \\ 
 & & PR & 3.26 & 1.00 & 112.96 \\ 
  & & CSIR & -0.02 & 0.94 & 0.72 \\ \hline
\end{tabular}
\label{tab:sim_results}
\end{table}

In all the simulations, cSIR gives results with small and stable bias close to 0. The other two methods estimate the SIR with bias frequently exceeding 0.5. Since the true SIR is between 1 and 2 in all of our simulations, this magnitude of bias is large. 

cSIR's 95\% confidence interval has a stable approximately 95\% coverage rate of the true SIR across all simulations. Due to high bias, CDC often gives low coverage of the true SIR, in some cases lower than 50\%. PR, on the other hand, is unstable due to the small number of samples from the exposed population and this instability leads to extraordinarily wide confidence intervals and, therefore, highly conservative coverage rates. 

 For cSIR, coverage of the null value decreases as the true SIR increases. We begin to see reasonable power to detect non-null SIRs when the true SIR is above 1.5, which means we are able to detect relatively small exposure effects. CDC's coverage of the null value is erratic and does not reflect trends in the true SIR, i.e., coverage of the null does not consistently decrease as the true SIR value increases. Above we discussed how PR's instability results in highly conservative coverage, and we see that this result holds not only for the true SIR but also for the null value. 
 
 In summary, the simulation results demonstrate that our approach, built on causal inference procedures, provides more reliable and stable results than existing alternative cancer cluster investigation procedures. We note that the structure of these simulations is favorable to CDC's method because we have assumed that the CDC's method is studying the true exposed population and time period, when in fact the CDC's protocol does not consider exposures when structuring the SIR analysis and therefore would be unlikely to analyze the appropriate population. Moreover, our simulation structure is also favorable to the Poisson regression method, as all the associations in the simulations are linear. In the presence of non-linear relationships, we expect even greater gains in our method compared to others, because matching automatically resolves even non-linear confounding effects. Yet, even in these conditions favorable to CDC and PR, the improvements offered by our proposed approach are clear.

\section{An investigation of kidney cancer incidence in Endicott, NY}

We define the boundaries of the exposed area in Endicott based on the work of previous NY state investigations, which determined the boundaries of the area affected by TCE vapor intrusion \citep{doh2006health}. We use in our analysis all CBGs fully or partially overlapping the exposed area. This leads to eight exposed CBGs, as shown in Figure~\ref{fig:maps}. TCE vapor exposure is only known to have been present in Endicott in 2002 and after, it is unclear when it began affecting the community; therefore, the appropriate time period to study is not obvious. Such uncertainties are likely to plague any cancer cluster investigation. The time period under study here, 2005-2009, was chosen primarily based on cancer data availability. Because urinary tract cancer latency periods are relatively short compared to other cancers \citep{yuan2010kidney}, any effects of TCE vapor exposure from the early 2000s or prior may be already be detectable during this time period.

\subsection{Data}
We use the SEER+ cancer incidence data described in Section~\ref{sss:data} for the years 2005-2009. For kidney/renal pelvis cancer incidence, we use all diagnoses in this time period with ICD-0-3 codes C649 and C659, and for bladder cancer incidence we use ICD-0-3 codes C670-C679. We obtained potential TCE exposure information for SEER+ areas from the EPA’s publicly available Toxics Release Inventory (TRI) data \citep{epa2018tri} and Superfund site data \citep{epa2018spf}. The use of over 650 toxic chemicals, including TCE, is tracked by the EPA. Businesses manufacturing or using more than a specified threshold amount of any one of these chemicals (and meeting certain other criteria) are required to submit yearly release reports to the EPA. Current and historic information about the location of these businesses, as well as the chemical types and amounts used by each, is provided to the public via the TRI data. The geocoded locations of all the Superfund hazardous waste sites, many of which have been contaminated by TCE, are also available through the EPA.

We employed the TRI and Superfund site location data to create a binary indicator of potential TCE exposure, around or before the time of Endicott's TCE vapor exposure, for each CBG in SEER+. We classify a CBG as potentially exposed to TCE if (a) a facility using TCE in or before 2000 or a Superfund site is/was located within its boundaries or (b) a facility using TCE in or before 2000 or a Superfund site is/was located within 2 miles of its centroid. We allow a CBG to serve as a potential matched control for the Endicott CBGs if it is classified as having no potential for exposure to TCE.

Finally, for each CBG in SEER+, we have collected data from ESRI Business Analyst \citep{esri2018} on the following potential confounders of the association between TCE exposure and cancer incidence (many overlap with those used to constuct simulations): percent of the population age 65+ (P65+), percent of the population male (PMale), percent of the population white (PWhite), rural indicator (Rural), percent of the adult population unemployed (Unemploy), average commute time (Commute), median household income (Income), total dollars spent on smoking products as a portion of per capita income (MoneySmoke), 
percent of total dollars spent on food that was spent on food outside the home (MoneyFood), percent of the population that reports exercising at least 2 times per week (Exercise), and percent of the population working in the agriculture, mining, construction, or manufacturing industries (Industry). A similar dataset could be constructed from US Census or American Community Survey data, if exclusively public data sources are desired. Because confounders should precede exposure, these confounder data come from the year 2000, just prior to the time that TCE vapor exposure was detected in Endicott.

\subsection{CBG cancer incidence prediction and cSIR estimation}
\begin{figure}[h]
\centering
\includegraphics[width=7in]{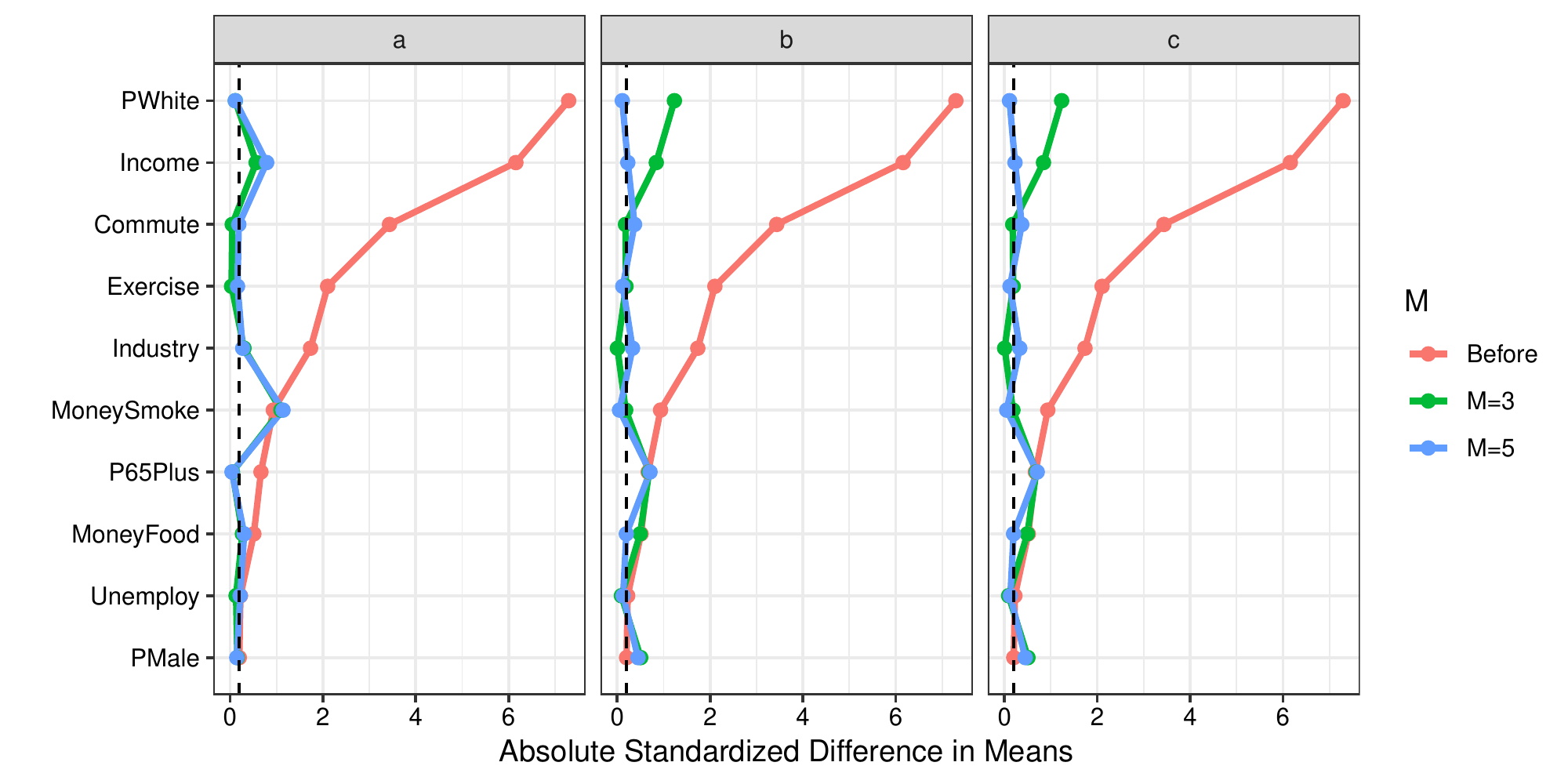}
\caption{Balance before matching and after 3:1 and 5:1 ratio matching on the mahalanobis distance (a), propensity scores estimated via linear model (b), and propensity scores estimated via generalized additive model. Covariates are considered well-balanced if the absolute standardized difference in means is less than 0.2 (marked by the dashed line).}
\label{fig:balance}
\end{figure}

R code for the analysis is available at \url{https://github.com/rachelnethery/causalSIR}. Our first step in estimating the cSIR is to identify matched control CBGs for the Endicott CBGs from the SEER+ data. We test different approaches to matching in search of a method that provides a reasonable compromise between our desire for (1) good confounder balance across exposure groups and (2) a substantial number of controls to stabilize the estimation. Both 3:1 and 5:1 nearest neighbor matching are applied to the data and for each ratio, 3 different distance metrics are tested: mahalanobis distance, distance in propensity scores estimated via linear model, and distance in propensity scores estimated via generalized additive model. Figure~\ref{fig:balance} shows the balance of each confounder before matching and after application of each of these matching methods (Rural, a binary variable, is not shown in the figure but is matched on exactly). The matching procedures dramatically improve the balance in most confounders. The different methods provide comparable results, and, for our analysis, we choose to use the matched data from the 5:1 matching on propensity scores estimated via linear model. Although the standardized differences in means after matching are not all less than the generally recommended (but arbitrary) threshold of 0.2, we are not concerned about these minor deviations from perfect balance, because we are also adjusting for the confounders in the analysis phase modeling.

\begin{figure}[h]
\centering
\includegraphics[scale=.6,page=1]{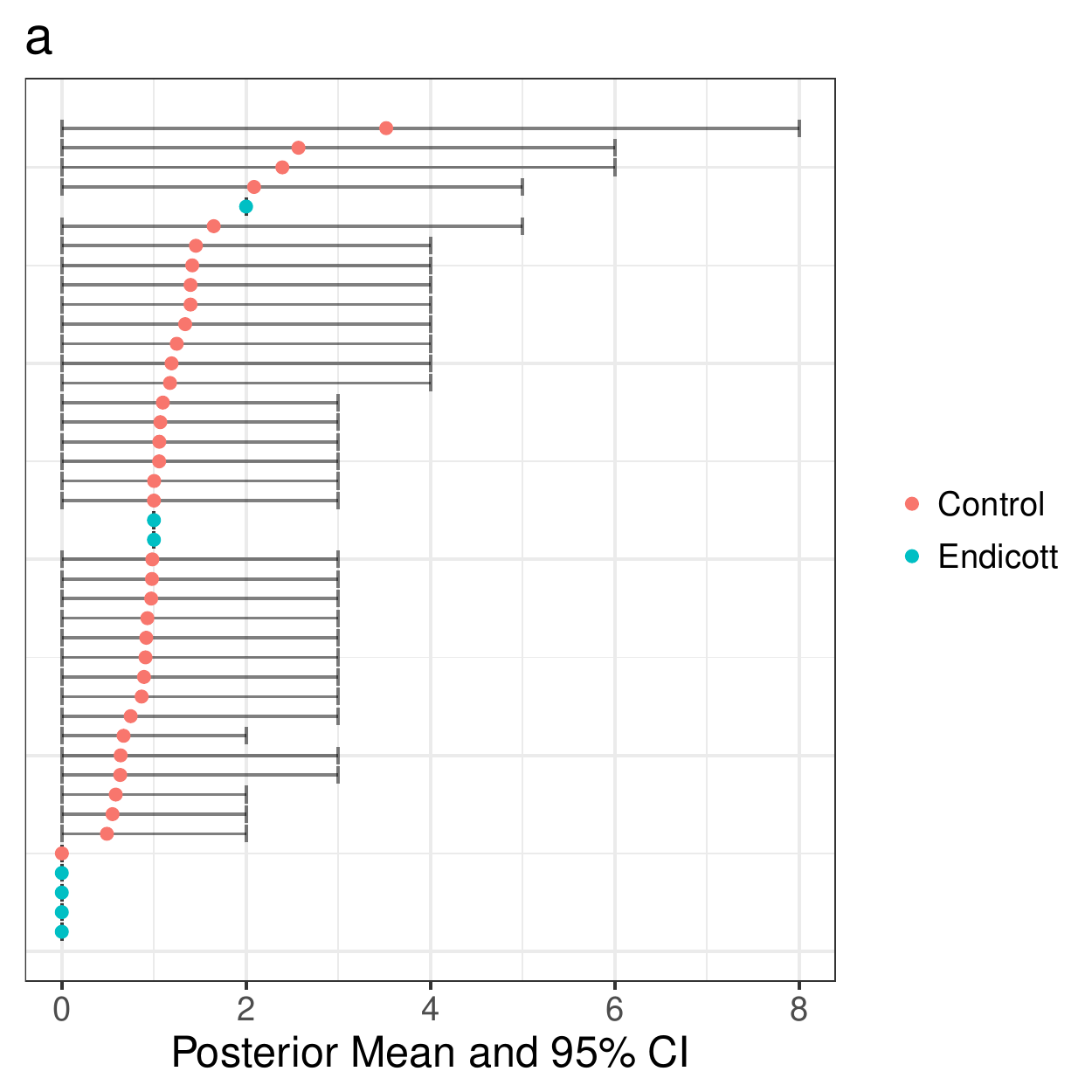}
\includegraphics[scale=.6,page=2]{endicott_preds.pdf}
\caption{Predicted kidney cancer incidence (a) and bladder cancer incidence (b) and 95\% credible intervals for matched control CBGs and observed incidences for Endicott CBGs.}
\label{fig:pred}
\end{figure}

Due to the spatial over-aggregation of the SEER data, the next step in the analysis is to apply the joint Bayesian model to predict the CBG kidney cancer and bladder cancer incidence for the matched controls and fit the loglinear model for cSIR estimation. We fit the prediction models using the CBG cancer incidence data from New York and the confounder variables described above as predictors, collecting 200,000 posterior samples. For kidney cancer, the prediction model parameter estimates reveal significant positive associations between each of the following variables and the proportion of the county's cancer incidence that falls into a given CBG: MoneySmoke, P65+, PMale, PWhite, Income, and Industry. A significant negative association with MoneyFood is detected. For bladder cancer, significant positive associations are found with MoneySmoke, Rural, P65+, PMale, PWhite, Income, and Industry. Significant negative associations are found with MoneyFood, Commute, and Exercise. We use these models to collect posterior predictive samples of the kidney and bladder cancer incidence for the matched control CBGs. Figure~\ref{fig:pred} shows the posterior means and 95\% credible intervals for the predicted incidences in the matched controls, compared to the observed incidences in the Endicott CBGs.

In the cSIR estimation models, all confounders are included besides Rural, because all CBGs in the matched dataset are non-rural. The resulting cSIR and 95\% credible interval for kidney cancer are 0.73 (0.28, 1.48). Using the other 5 matching methods/ratios considered above, the kidney cancer cSIR estimates range from 0.53 to 0.71. None are statistically significant. We also applied the CDC's SIR analysis as described in Section~\ref{ss:simmethods} to estimate the SIR for kidney cancer in the TCE-exposed area of Endicott, using all of NY state as the background population to compute the expected kidney cancer incidence. The resulting SIR and 95\% confidence interval are 0.73 (0.24, 2.02). The similarity in the SIR and cSIR results may mean that confounding has little effect on the kidney cancer analysis.

For bladder cancer, the cSIR and 95\% credible interval are 1.38 (0.72, 2.35). The estimates produced by the other 5 matching methods/ratios range from 1.30 to 1.38, with none statistically significant. Using the CDC's method, the bladder cancer SIR estimate and 95\% confidence interval are 2.05 (1.17, 3.33). The statistically significant results of the SIR analysis could be due to lack of adjustment for confounding and should be used with caution.

We also perform a sensitivity analysis in which we omit MoneyFood, MoneySmoke, and Exercise from the matching/adjustment set. The remaining variables represent the subset that can be easily obtained for small areas from the census or other publicly available sources. For both the kidney and bladder cancer models, we observe only minor changes in the results using this subset of the confounders. Using 5:1 propensity score matching, the kidney and bladder cancer cSIR estimates and 95\% credible intervals are 0.67 (0.24, 1.40) and 1.27 (0.65, 2.23), respectively.

If we are willing to make the assumptions of SUTVA, ignorability, and causal consistency described in Section~\ref{ss:csirassumptions}, then our results are endowed with a causal interpretation. However, the assumption of no unobserved confounding, which is untestable, could be violated in this analysis. As in most studies, we do not have information about potentially important confounders such as diet, accessibility of health care, and exposure to other sources of pollution/contamination. To interpret our results as causal, we must assume that these factors are not confounders of the TCE exposure-kidney/bladder cancer relationship.

\section{Discussion}
In this paper, we have introduced a causal inference framework for cancer cluster analyses, which relies on a priori identification of sources of hazard that could cause increased cancer incidence. By constructing statistical analyses around exposure hypotheses rather than observed cancer outcomes, the silent multiple comparisons problem associated with the traditional approach to cancer cluster investigations is resolved so that statistically valid results are possible. Moreover, this approach allows us to directly ask and answer the question of interest-- whether exposure to a specific hazard caused increased cancer incidence in a community.

Using the potential outcomes framework, we develop a causal analog of the standardized incidence ratio typically used to evaluate cancer clusters, the causal SIR, and provide identifying assumptions. We propose a two-stage Bayesian model that resolves the problem of spatial over-aggregation in cancer incidence data. This model, applied to a matched dataset, allows the cSIR can be estimated from publicly available data. In simulations, our statistical approach was shown to provide dramatically improved results, i.e., less bias and better coverage, than the current approach to SIR analyses. Finally, we demonstrated the use of our method by applying it to investigate whether TCE vapor exposure, resulting from a chemical spill dating back the the 1970s, caused increased kidney or bladder cancer incidence in Endicott, NY during 2005-2009. Our method did not produce any statistically significant results. The CDC's SIR analysis method found a significant result for bladder cancer, likely attributable to unresolved confounding effects.

A possible concern about this causal inference approach to cancer cluster analyses is related to the appropriateness of the SUTVA assumptions in this setting. One requirement of SUTVA is that there is only one ``version'' of exposure. In many investigations, some CBGs in the exposed area will be more highly exposed than others, leading to multiple different versions of exposure. Another requirement of SUTVA, referred to as ``no intereference'', is that the treatment status of one unit cannot affect the outcome of another unit. This assumption also may be dubious in many cancer cluster analyses, because within the exposed community, some people may live in one exposed CBG and work in another exposed CBG. Chronic exposure in ones workplace may be just as likely to cause cancer as exposure in the home; therefore, the exposure status of the workplace CBG may impact the cancer incidence in the CBG of residence. This would be a violation of SUTVA. In such a setting, our method can still be used to evaluate associations between exposure and cancer incidence, but the results may not be causal. In future work, extensions of methods to handle SUTVA violations in estimation of average causal effects could be integrated into the causal SIR framework \citep{vanderweele2013causal,barkley2017causal,papadogeorgou2017causal}.

While our method provides an improvement over existing methods for cancer cluster investigation, it has numerous limitations. First, while our method resolves the silent multiple comparisons problem and provides rigorous adjustment for confounding, other complicated issues that affect all cancer cluster investigations such as population migration are not directly addressed by this method. If substantial population migration has occurred in the community under study between the time of exposure and the time that cancer outcomes are investigated, then the results from this method may not be reliable and should not be interpreted as causal. Moreover, these methods need more testing and development in the setting of very rare cancer with many zeros counts in small areas. As we discuss in Section~\ref{s:methods}, our method for addressing spatial over-aggregation relies on a Poisson likelihood and must be extended to handle zero-inflation and extra-Poisson variation. Bayesian approaches to modeling zero-inflated data have been studied at length \citep{ghosh2006bayesian,ozmen2010bayesian,liu2012bayesian}, and these ideas could be integrated with our joint modeling approach. Additionally, more work is needed to adapt this framework to the setting in which multiple exposures affecting a community may have synergistic effects on cancer incidence. In some contexts, small sample size of the matched data and a potentially large number of confounders may mean that causal methods for $p>N$ need to be integrated into this approach. 

Likely the most challenging aspects of applying these methods in real cancer cluster investigations will be (1) determining the population and time period exposed to a given source and (2) collecting reliable data. With regards to the former, we remark that the exposure hypotheses on which analyses are based do not have to be perfect nor unanimously agreed upon. First, multiple different potential exposures can be considered and analyzed (separately), i.e., a single exposure for investigation need not be settled on from the beginning. Moreover, while some research should be done regarding the area, time period, and cancer types reasonably associated with a given exposure, these determinations need not be set in stone in order to proceed with statistical analyses. Different reasonable specifications of population, time period, and cancer types could be tested and the results multiple comparisons adjusted accordingly, using standard multiple comparisons corrections like Bonferroni.

Obtaining reliable data to carry out these analyses is a less forgiving endeavor. While a good deal of confounder data is readily available from the census, cancer incidence and exposure data are more limited. As described here, a few states are beginning to take the lead in public release of small area cancer incidence data. If this movement spreads, it stands to deliver huge improvements to the efficiency and reliability of cancer cluster investigations. Exposure data may be difficult to obtain for certain types of hazard, and its reliability is often dubious. For instance, the TRI data only represent businesses using large amounts of certain chemicals, and businesses self-report usage to the TRI database. Moreover, the TRI data do not capture events like spills of chemicals that may put communities at highest risk. In order to carry out cancer cluster investigations with maximal rigor, more work is needed both to collect better data and to make the data more easily accessible.

\section{Acknowledgements}
Support for this work was provided by NIH grants R01GM111339, R35CA197449, R01ES026217, P50MD010428, DP2MD012722, R01ES028033, and R01MD012769. HEI grant 4953-RFA14-3/16-4 and EPA grant 83615601 also funded this work.

\bibliographystyle{chicago}
\bibliography{csir}

\end{document}